\documentclass[a4paper,fleqn,usenatbib]{mnras}

\usepackage{mathptmx}

\usepackage[T1]{fontenc}
\usepackage{ae,aecompl}
\usepackage{graphicx}
\usepackage{epstopdf}

\def\ie{{\it i.e., }}
\def\be{\begin{equation}}
\def\ee{\end{equation}}
\def\bea{\begin{eqnarray}}
\def\eea{\end{eqnarray}}
\newcommand{\eg}{{\it e.g., }}

\title[Measuring metallicity from photometry with ML]{How to measure metallicity from five-band photometry \\ with supervised machine learning algorithms}

\author[V. Acquaviva]{Viviana~Acquaviva, $^1$\thanks{vacquaviva@citytech.cuny.edu}  \\
$^1$ Department of Physics, CUNY NYC College of Technology, 300 Jay Street, Brooklyn NY 11201\\
}

\date{Accepted 2015 November 16. Received 2015 November 16; in original form 2015 September 12}
\pubyear{2016}

\begin{document}

\label{firstpage}
\pagerange{\pageref{firstpage}--\pageref{lastpage}}
\maketitle
    
\begin{abstract}

We demonstrate that it is possible to measure metallicity from the SDSS five-band photometry to better than 0.1 dex using supervised machine learning algorithms. Using spectroscopic estimates of metallicity as ground truth, we build, optimize and train several estimators to predict metallicity. We use the observed photometry, as well as derived quantities such as stellar mass and photometric redshift, as features, and we build two sample data sets at median redshifts of 0.103 and 0.218 and median r-band magnitude of 17.5 and 18.3 respectively. We find that ensemble methods, such as Random Forests of Trees and Extremely Randomized Trees, and Support Vector Machines all perform comparably well and can measure metallicity with a Root Mean Square Error (RMSE) of 0.081 and 0.090 for the two data sets when all objects are included. The fraction of outliers (objects for which |Z$_{\rm true}$ - Z$_{\rm pred}$| $>$ 0.2 dex) is 2.2 and 3.9\% respectively, and the RMSE decreases to 0.068 and 0.069 if those objects are excluded. Because of the ability of these algorithms to capture complex relationships between data and target, our technique performs better than previously proposed methods that sought to fit metallicity using an analytic fitting formula, and has 3x more constraining power than SED fitting-based methods. Additionally, this method is extremely forgiving of contamination in the training set, and can be used with very satisfactory results for sample sizes of a few hundred objects. We distribute all the routines to reproduce our results and apply them to other data sets.
\end{abstract}


\section*{Introduction}

The existence of a tight correlation between the stellar mass and metallicity of galaxies is a well-established evidence in galaxy evolution (e.g., \citealt{Tremonti2004}). More recently, it has been proposed that this correlation is the result of a more fundamental relation among metallicity, star formation rate, and stellar mass (\eg \citealt{LaraLopez2010,Mannucci2010}). The existence of this so-called fundamental metallicity relation (FMR) is still controversial, and a deeper understanding of whether this relation exists and how it evolves with redshift would provide insight into the fundamental mechanisms that regulate growth and star formation in galaxies. In fact, the abundance of metals in galaxies (defined throughout this paper as the oxygen-to-hydrogen abundance) is determined by the stellar mass of the galaxy, the amount of inflows and outflows that can dilute the metal content, and the gas mass of the galaxy, which also depends crucially on the galaxy's outflows and merger history (\eg \citealt{Dave2012}). 
So far, the investigation of the existence and evolution of the FMR has often relied on spectroscopic measurements of the strength of emission lines used as estimators of metallicity, such [Ne III] to [O II], [O III] to [O II], and R23 (([O III] + [O II])/ H$\beta$). As a result, it has been difficult to extend these studies to high-redshift, low-stellar mass objects samples, for which the amount of available data remains limited (\eg \citealt{DeLosReyes2014} and references therein). 
A major game changer would come from the opportunity to extend the study of the mass-metallicity relation and FMR to larger samples at high redshift and low stellar masses by measuring metallicity from photometric data. Traditional SED fitting methods are promising \citep{Dye2008, Pacifici2012}, but sampling the likelihood as a function of metallicity is difficult because of the limited number of available templates, and the fact that the dependence of the SED on metallicity is highly non-linear. Furthermore, model-based SED fitting constrains the stellar metallicity, rather than the gas-phase metallicity which enters the FMR relation. More recently, it has been recognized that the mass-metallicity correlation can be tightened by also considering luminosity and rest-frame colors \citep{Sanders2013}. In this paper, we propose to use supervised machine learning algorithms to optimally investigate the correlation between these quantities, and we demonstrate that if a moderate-size, unbiased spectroscopic calibration set is available, it is possible to measure metallicity to better than 0.1 dex precision with the five-band SDSS photometry. 

The paper is organized as follows. In Sec. \ref{sec:ML} we give a brief introduction about supervised machine learning techniques and describe the algorithms that we propose to use. In Sec. \ref{sec:optimize} we describe our optimization process, which consists of comparing and selecting different estimators, as well as of data cleaning and feature selection and engineering. In Sec. \ref{sec:results} we apply the optimized algorithm to two sample data sets within the SDSS catalog, measuring the metallicity of SDSS galaxies and its uncertainty, and we compare our results to the current literature. In Sec. \ref{sec:spectrophoto} we combine photometric data with spectroscopic measurements of emission lines fluxes and investigate the improvement in the determination of metallicity awarded by each one of them. In Sec. \ref{sec:applications}
we consider the applicability of our method to smaller samples, and we forecast the improvement in the metallicity measurement for the LSST main survey. Sec. \ref{sec:conclusions} summarizes our findings.

\begin{figure*}
\begin{center}
\includegraphics[width=0.33\linewidth]{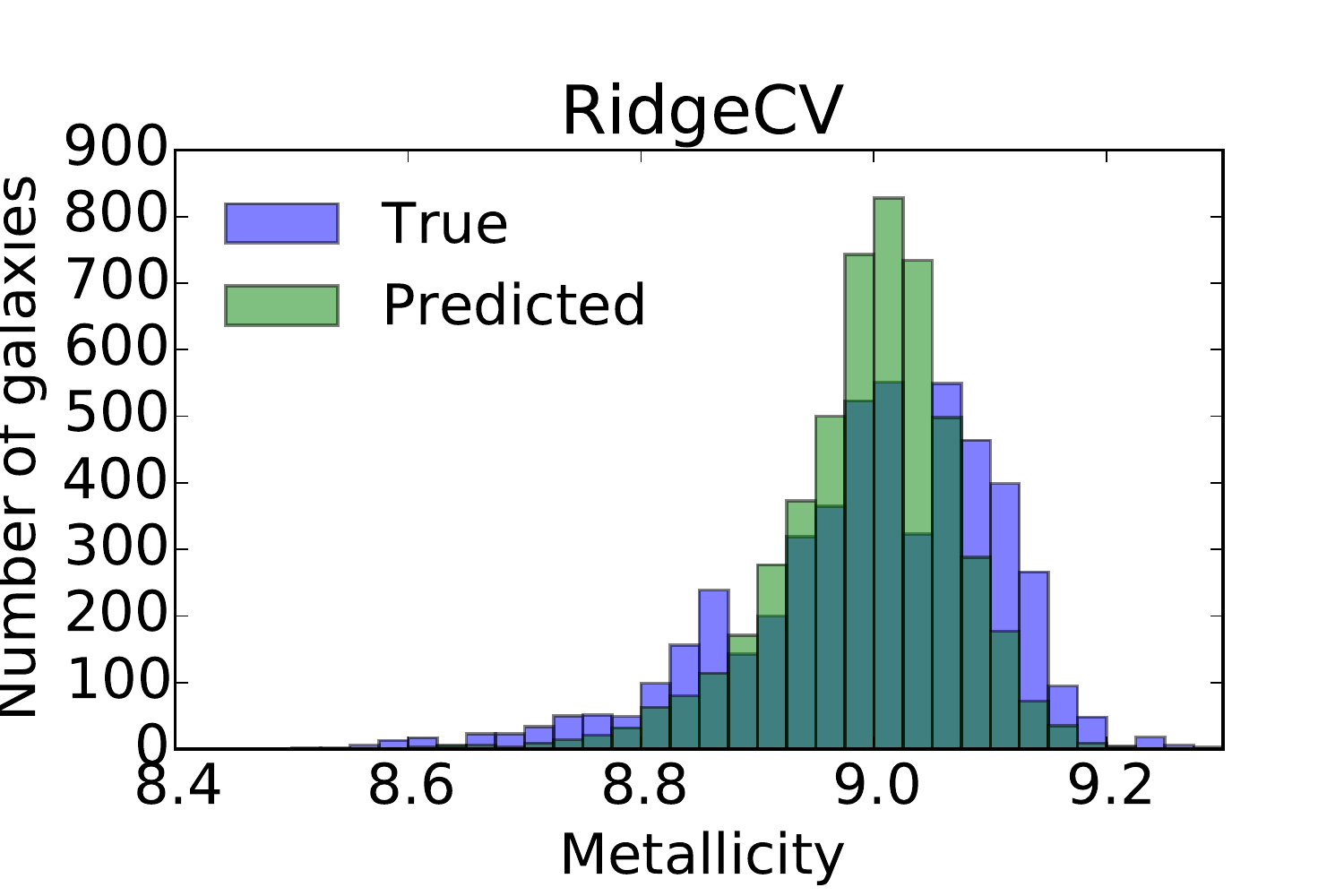} 
\includegraphics[width=0.33\linewidth]{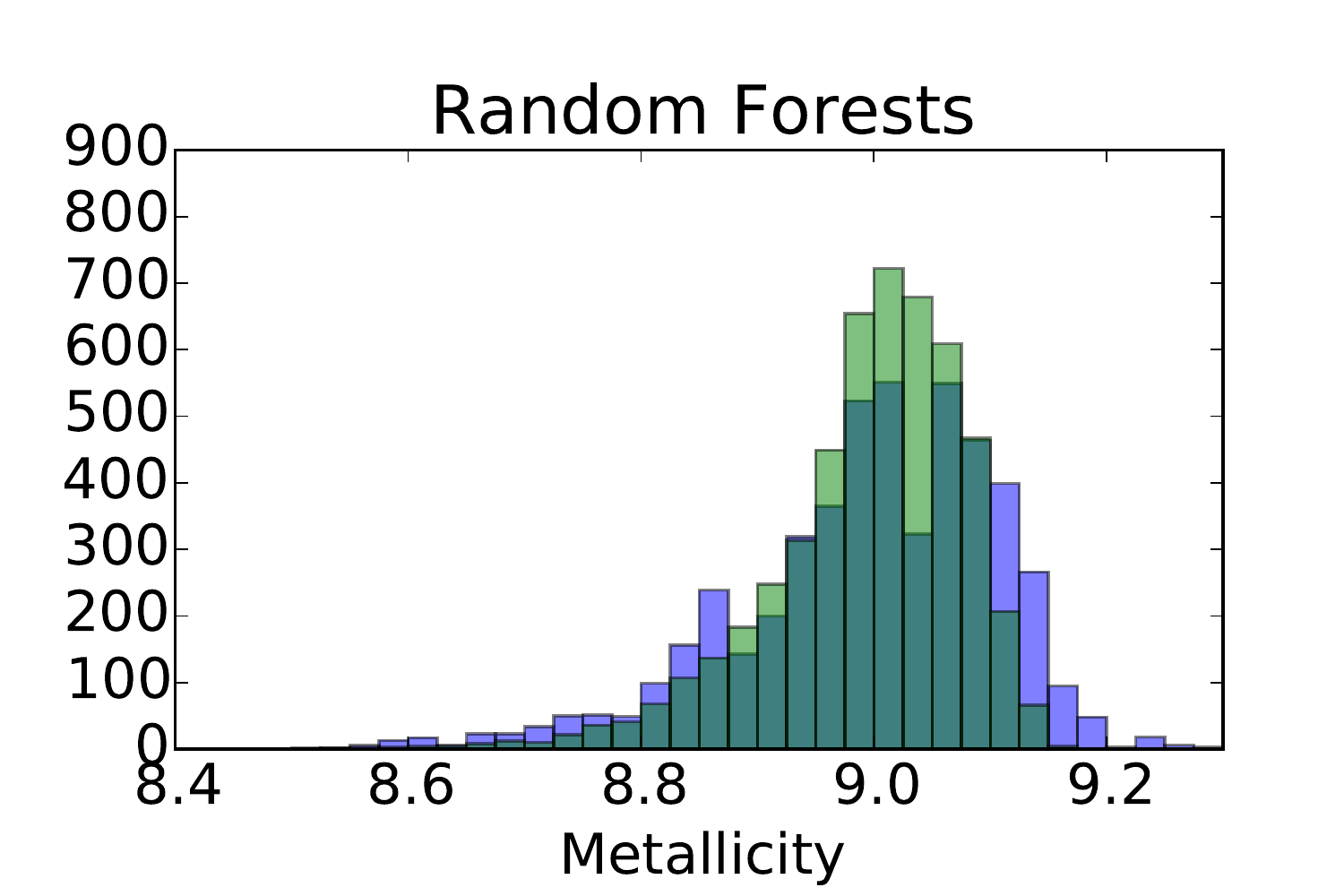}
\includegraphics[width=0.33\linewidth]{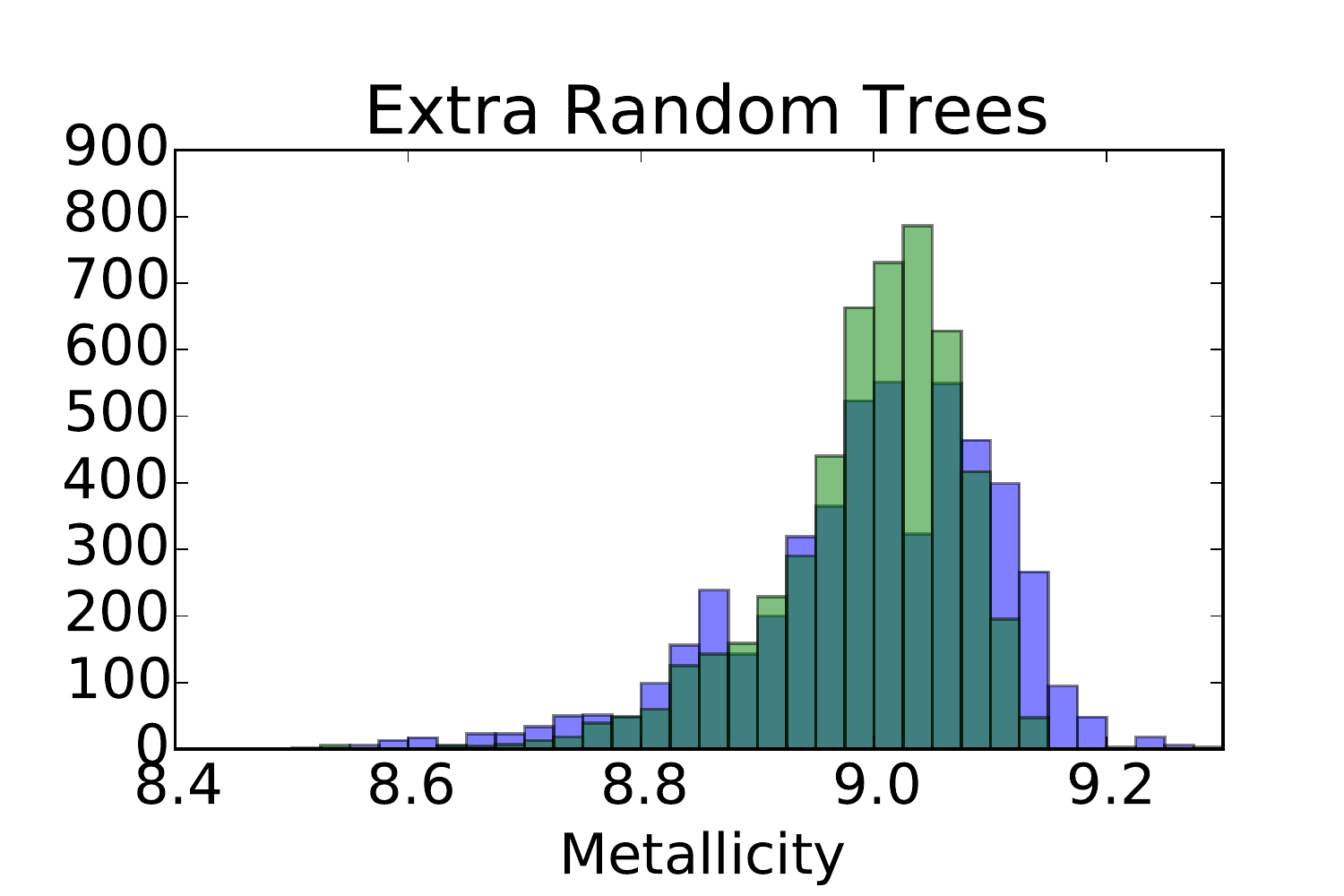} \\
\includegraphics[width=0.33\linewidth]{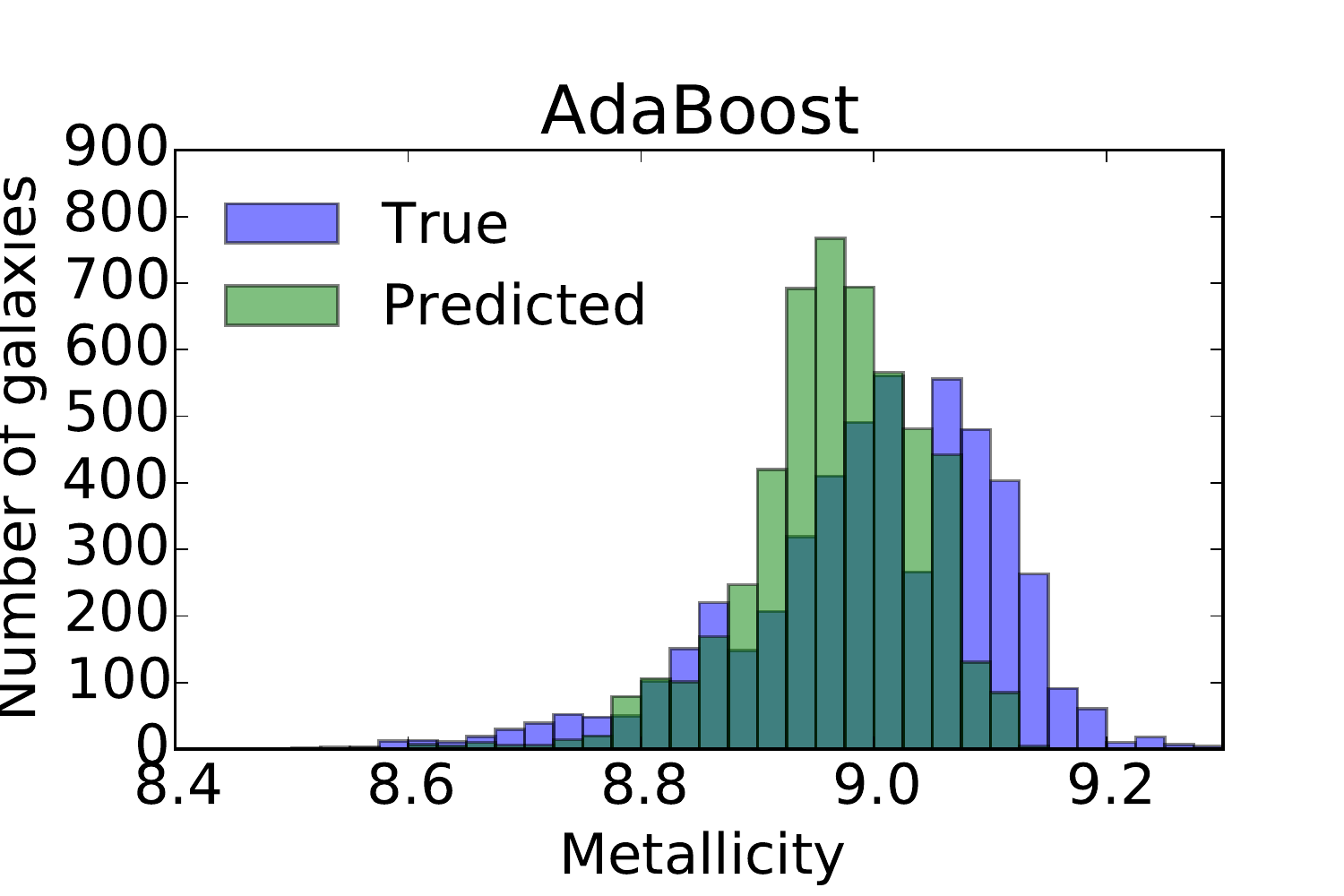}
\includegraphics[width=0.33\linewidth]{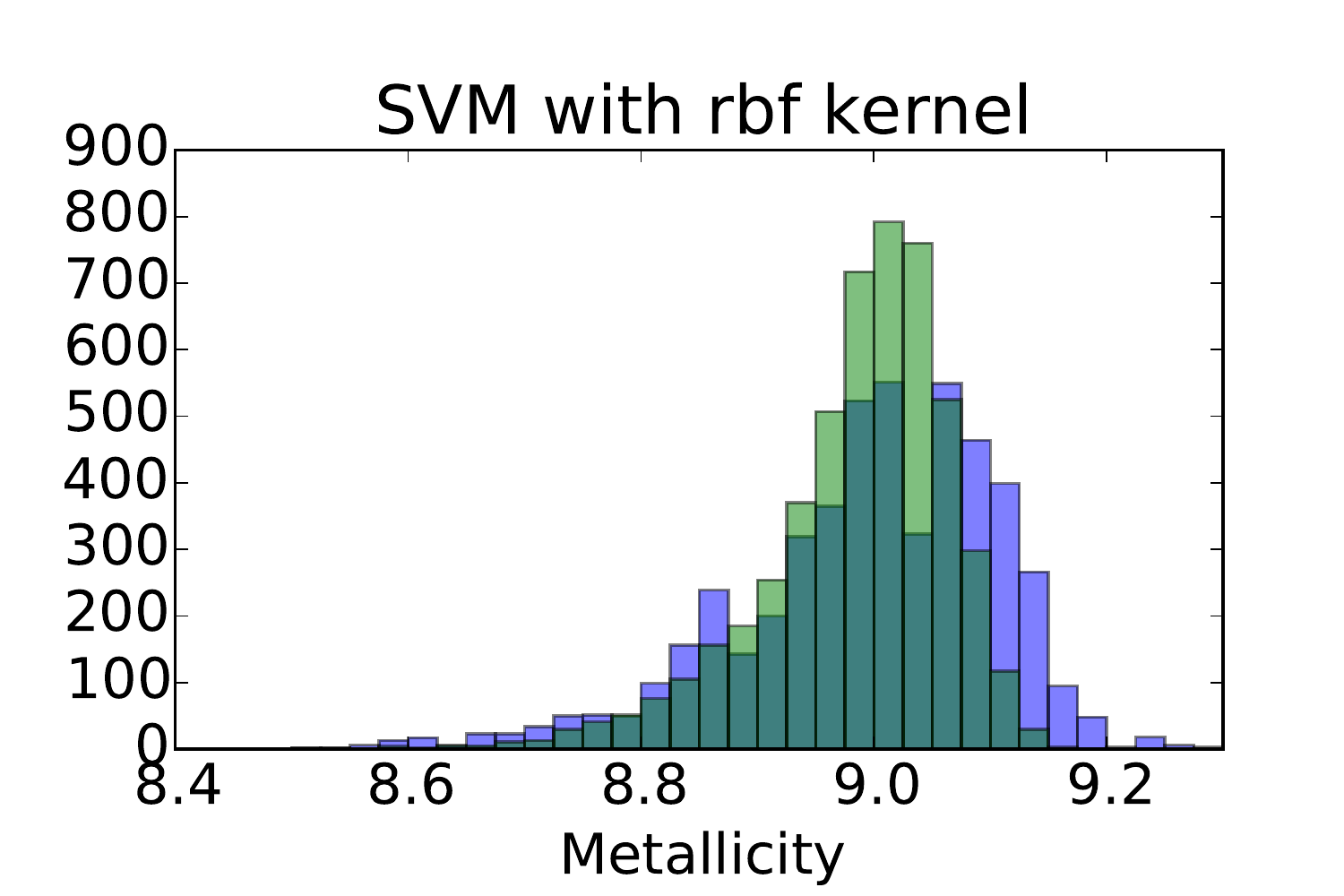}
\caption{Normalized distribution of ``true" and predicted metallicity values for the five algorithms we consider and the first test set described in Sec. \ref{subsec:Opti}, composed by $\sim$ 5,000 objects at $0.09 < z < 0.12$. Random Forests, Extremely Randomized Trees and Support Vector Machines exhibit the lowest bias.}
\label{fig:distributions}
\end{center}
\end{figure*}

\section{Supervised machine learning methods}
\label{sec:ML}

Machine learning (ML) is a set of tools used to infer a relation between known variables (either observable quantities, or an engineered combination of them) and unknown variables, which we desire to determine - learn, in ML jargon. The known quantities are called {\it features} and the unknown ones are called {\it target}. In supervised learning, this relation is inferred by means of a training set, which is a subset of the data for which both the features' and the target's values are known. The training set can be split into a cross-validation set, which is used to tune the parameters of the learning algorithm until the optimal performance is reached, and a test set, which is used to evaluate the expected performance of the algorithm on a ``new" set of data that never took part in the training process. The performance achieved on the test set can be used as a metric to select the best algorithm. 

In this paper, we compare five different learning algorithms: regularized ridge regression, random forests of trees (RF), extremely randomized trees (ERT), boosted decision trees (AdaBoost), and support vector machines (SVM). The reason behind these choices are the following. First of all, the well-known paper \cite{Caruana05anempirical} showed that for eleven different supervised learning problems and ten supervised learning algorithms, the latter four methods have a $95\%$ combined probability of being the best-performing estimator. Second, these algorithms have very different responses to the two most common problems in machine learning: over-fitting or high variance (excessive tailoring of the algorithm to the training set), and under-fitting or high bias (failure to capture the most important features and achieve satisfactory performance). In approaching a new problem with no prior information about the possibility of high bias or high variance, these algorithms span the range of possible solutions. Finally, we included a simple logistic regression algorithm, which is not expected to perform as well as the others unless the learning rule is simple, but has a much faster time scaling. All algorithms are implemented using the scikit-learn package in Python \citep{scikit-learn}.

\begin{table*}
\begin{center}
\resizebox{\linewidth}{!}{
\begin{tabular}{|c|cccc|}
\hline
&&&&\\
Algorithm & Parameter & Range & Optimal, test set 1 & Optimal, test set 2  \\
&&&&\\
\hline
&&&&\\
Ridge Regression & Regularization $\alpha$ & [0.1, 1.0, 10] & 1.0 & 10 \\
&&&&\\
\hline
& Number of Estimators & [10, 20, 40] & 40 & 40  \\
Random Forests & Min samples to split a branch & [2, 4, 6, 8] & 2 & 4  \\
& Min samples in a leaf & [2, 4, 6, 8] & 8 & 4   \\
\hline
& Number of Estimators & [10, 50, 100] & 100 &  50 \\
Extremely Random Trees & Min samples to split a branch & [2, 4, 6, 8] &  2 &  4 \\
& Min samples in a leaf & [2, 4, 6, 8] & 8 & 4  \\
\hline
& Number of Estimators & [10, 50, 100] & 50 & 50  \\
AdaBoost & Loss function & [linear, square, exp] & exp & exp  \\
  & Max depth in weak estimator tree & [4, 6, 8, 10] & 8 & 8  \\
\hline
  & kernel & [linear, rbf] & rbf &  rbf \\
 SVM & C (penalty function) & [1, 10, 100] & 1 & 10  \\
  & gamma (complexity of boundary) & [0.01, 0.1, 0.5] & 0.1 & 0.01  \\
\hline
\end{tabular}}
\caption{Parameter grid and optimal choices for the five algorithms we considered. For Random Forests, while adding more trees is in general beneficial, we verified that we had reached the "plateau" in performance at 40 trees; the improvement is less than half the standard deviation between 40 and 80 trees, despite a two-fold increase in CPU time.}
\label{tab:opti}
\end{center}
\end{table*}

\section{Building the optimal estimator}
\label{sec:optimize}

\subsection{Feature selection and engineering}

A critical ``human input" in optimizing machine learning algorithms is the selection of the features that are likely to carry the highest amount of information. 
This process also includes engineering features, \ie combining them in a smart way. For example, if we wanted to build an algorithm to ``learn" how much gas is needed to drive from point A to point B as a function of their geographical coordinates, adding a ``distance" feature would dramatically improve performance. In our case, the observable quantities are the five SDSS photometric bands (u, g, r, i, and z), and the target quantity is metallicity. We assume that the ground truth for metallicity is given by the estimates (based on spectroscopic data) by \cite{Tremonti2004}, which of course might differ from the ``true", physical gas-phase metallicity of these galaxies if such estimates are not accurate. Some additional features are also available: we include stellar masses and photometric redshifts, since we expect them to be highly correlated with metallicity, and they can be derived for similar data sets through SED fitting. Spectroscopic measurements of emission line strength are also likely to carry a good amount of information, but we do not include them in our nominal setup because we focus on purely photometric measurements. However, they will be discussed in Sec. \ref{sec:spectrophoto}. 
 
The five measurements of photometric brightness can also be combined in colors, as well as in colors raised to some power. Previous studies that tried to isolate correlations between colors and other quantities usually applied a K-correction to calculate rest-frame colors, but since redshift is one of the features, it is not necessary to do that here as long as the data are binned in relatively thin ($\delta z \sim 0.1$) redshift slices. We add the ten independent colors and the ten independent squared colors to our list of features. A similar approach was also used by \cite{Mannucci2010} and \cite{Sanders2013}. While machine learning algorithms (for example, SVMs with polynomial kernel) are often able to combine features in polynomial-like fashion, there is a great computational efficiency advantage in specifying combinations of features explicitly. Furthermore, it is often convenient to start with an inclusive list of features, rank them in order of their importance for the estimation of the target, and if necessary eliminate features that don't add any information or introduce excessive noise. We further discuss this issue in Sec. \ref{subsec:Diag}.

\begin{table*}
\begin{center}

\resizebox{\linewidth}{!}{
\begin{tabular}{|c|cccccc|}
\hline
Test set & Algorithm & RMSE & RMSE& OLF & r2 score & CPU time \\
 & & (all objects)  &   (no outliers)  & & & to fit training data \\
\hline
& Ridge Regression & 1 & 1 & 1 & 1 & 1 \\
& Random Forests & 0.96 & 0.98 & 0.9 & 1.04 & 242  \\
 0.09 $<$ z $<$ 0.12& {\bf Extremely Random Trees} & 0.96 & 0.98 & 0.89 & 1.05 & 80\\
& AdaBoost & 1.0 & 1.07 & 0.8 & 0.9 & 173  \\
& SVM (rbf kernel) & 0.95 & 0.99 & 0.8 & 1.03 & 105 \\
\hline
& Ridge Regression & 1 & 1 & 1 & 1 & 1\\
& Random Forests  & 0.94 & 0.93 & 0.88 & 1.15 & 311  \\
 0.2 $<$ z $<$ 0.25& {\bf Extremely Random Trees} & 0.91 & 0.92 & 0.75 & 1.16 & 69 \\
& AdaBoost & 0.94 & 0.95 & 0.72 & 1.12 & 301 \\
& SVM (rbf kernel) & 0.99 & 0.97 & 1.0 & 1.03 & 48 \\
\hline
\end{tabular}}
\caption{A comparison of the five optimized algorithms for two test sets, at lower and higher redshift. Results are normalized around the performance of the ridge regression algorithms. Low values are better for all metrics except $r_2$ scores, for which increasing values indicate increasing resemblance between ground truth and prediction. Extremely Randomized Trees have the best performance overall, trailed by Random Forests, which are considerably slower, and SVMs, which perform worse in the second test set. }
\label{Tab:Comparison}
\end{center}
\end{table*}
\vspace{1cm}

\begin{table*}
\begin{center}
\resizebox{\linewidth}{!}{
\begin{tabular}{|c|ccccccc|}
\hline
 & Nominal & No AGN flag & No magnitude cut & No dust correction & Ignoring redshift flag & Using C model mags & Scaling \\
\hline
Test set 1, RMSE & 1.0 & 1.0 & 1.0 & 0.98 & 0.99 &  1.01 & 1.0  \\
Test set 1, OLF  & 1.0 & 1.03 & 1.4 & 0.95 & 0.95 & 1.02 & 1.0  \\ 
Test set 2, RMSE & 1.0 & 1.01 & 0.99 & 1.0 & 0.99 & 1.01 & 1.03  \\
Test set 2, OLF & 1.0 & 1.0 & 1.05 & 1.02 & 0.97 & 1.0 & 1.05  \\ 
\hline
\end{tabular}}
\end{center}
\caption{Variation in performance metrics as a result of using different options in cleaning and organizing data.}
\label{Tab:Diagnostics}
\end{table*}

\subsection{Algorithm optimization and selection}
\label{subsec:Opti}

The five algorithms we selected as possible candidates are Ridge logistic regression, Random Forests of trees, Extremely Randomized Trees, AdaBoost with decision trees, and Support Vector Machines. Ridge regression is a linear algorithm that looks to minimize the squared sum of the distance between model and data, with a built-in regularization procedure that effectively bounds from above the value of the coefficients of the linear fit. Random Forests, Extremely Randomized Trees and AdaBoost are all ensemble methods based on decision trees. Decision trees can be thought of as a flow chart where the path along the forks (the branches of the trees) is decided by the value of the features.
In Random Forests, different decision trees are built on random subsets of the data, tree splits are picked as the best splits among a random subset of the features, and the final outcome is determined as the average of the outcomes of all the trees. This method is particularly suitable for high-variance problems because the randomized regressor is unlikely to overfit the data. Extremely Randomized Trees go one step further in the randomization, by using random (rather than optimal) thresholds as the splitting rules for different branches of the trees.
Boosting algorithms, such as AdaBoost, are also based on building many models (in this case decision trees), and then combining to obtain a stronger model, but in this case the highest-performance trees are given higher weight (``boost") than the weaker ones, and thus the random nature of the regressor is less prominent and the method might be sensitive to noisy data and/or outliers. Finally, in Support Vector Machines, the feature space is transformed to a much higher dimensional space where the data are distributed more sparsely by means of a kernel function, and the mapping between input and output happens in this transformed space. SVMs are known to be accurate because there is great flexibility in selecting the kernel function and the algorithm might be able to select complicated combinations of the features, making them suitable for high-bias problem, but are slower and might be susceptible to over-fitting. 

To optimize the algorithms, we selected two sample data sets. We started by considering all objects in the SDSS main sample with metallicity measurements from \cite{Tremonti2004}, and for which a mass measurement was available.  We utilized the ``model" magnitudes rather than the ``C" magnitudes provided by the SDSS-JHU team\footnote{http://www.mpa-garching.mpg.de/SDSS/DR7/}, since they indicate that model magnitudes might lead to more accurate color estimation. We excluded objects with an ``AGN" flag and with a ``redshift quality warning" flag, and we applied a correction for galactic dust according to the maps of \cite{SFD98}.  For the first data set, we required objects to have a redshift between 0.09 and 0.12, and r-band magnitude lower than 18.0. This resulted in a sample of 25,316 objects with average/median r-band magnitude equal to 14.2/17.5, which we divided in a training set (80\%) and test set (20\%). For the second data set, we selected objects with redshift between
0.2 and 0.25, and r-band magnitude between 15.0 and 25.0. The resulting sample had 3,050 objects with average/median r-band magnitude of 18.3, which again we divided in a training set (80\%) and test set (20\%).

 We then tuned each of the five algorithms' hyper parameters by running a grid search with three-fold cross validation on the training set, selecting the best combination of parameters. The main metric that we use to evaluate the algorithms is the root mean square error (RMSE), which is defined as the average square of the difference vector between ground truth and prediction. The details of the optimized parameters can be found in Table \ref{tab:opti}.

The results of this comparison are reported in Table \ref{Tab:Comparison}. We report: the RMSE, the fraction of outliers (OLF), defined as the fraction of objects for which the true value is more than 0.2 dex away from the predicted value, the ``r$_2$ score", or {\it coefficient of determination}, which is a measure of the difference between predicted and true values, and is defined as 1 - (variance of data)/(mean of squared residuals), and the CPU time to fit the training data.

We use the Ridge regression algorithm as a benchmark and report the results for the other algorithms as a ratio with respect to the benchmark. Fig. \ref{fig:distributions} also shows the distribution of the predicted vs. true values for the five algorithms for the first of the two test sets. In both cases, we find that Random Forests, Extremely Randomized Trees, and Support Vector Machines have comparably good performances, with ERT being the absolute winner once time scaling is also taking into account. We will be using the ERT with the parameters described in Table \ref{tab:opti} in the following sections.

\subsection{Cleaning and processing data}

\label{Sec:DataCleaning}

One of the expected advantages of machine learning versus explicit methods is the ability to recognize and deal with outliers, so it is interesting to check if and by how much each of the selection criteria described in the previous section matters for the accuracy of the metallicity estimation. For this test we used the Extremely Randomized Trees algorithm, and we report results as ratios to the nominal setup.

To perform this test, we changed one of the selection criteria at the time for each of the sample data sets. This included ignoring the AGN and redshift quality flag, using ``C" magnitudes rather than model magnitudes, and forfeiting the magnitude cuts and the dust correction. We also scaled the features in the data set, by re-normalizing each feature to have zero mean and unit variance; this is common practice in machine learning since some algorithms might be biased if features have very different numerical ranges, although in our case the range of variation of all variables is within one-two orders of magnitudes, and the ERT algorithm is not expected to require scaling (unlike, for example, SVMs).  Results are reported in Table \ref{Tab:Diagnostics}. Overall, none of these factors affected the results at more than 1-$\sigma$ significance, indicating that the performance of supervised ML algorithms is quite robust to different choices in data selection and cleaning.

\section{Results}
\label{sec:results}

On the basis of the procedure described in the previous sections, we now describe the results for our optimal estimator (Extremely Randomized Trees) for the two test sets we have used. 

\begin{figure}
\includegraphics[width = \linewidth]{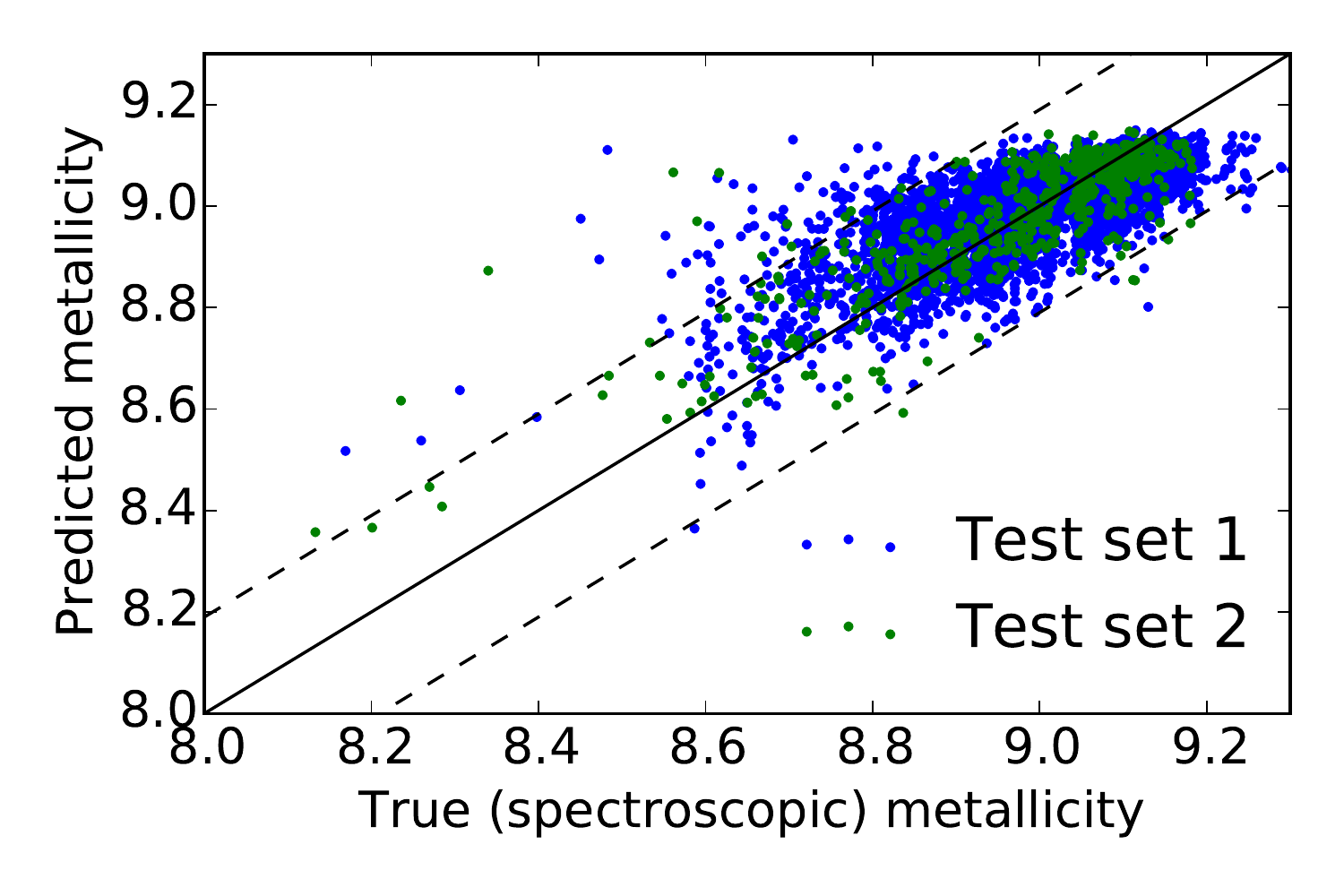}
\caption{Spectroscopic vs photometric metallicity for the two test sets described in the test. The solid line indicates 1:1 correspondence, and the dashed lines enclose the $\pm$ 0.2 dex deviation; objects outside these lines are classified as outliers. All objects in the test sets are included in the diagram. The fraction of outliers is 2.2\% for test set 1 and 3.9\% for test set 2.}
\label{Fig:BestResults}
\end{figure}

We ran our algorithm on two sets of features: at first, using only the information contained in the photometry (\ie the observed magnitudes and colors), without using any derived quantity such as stellar mass or photometric redshifts, and subsequently after including also these two features. We found that in both cases and for both test sets, we are able to measure metallicity with remarkable precision. For test set 1, the Root Mean Square Error is 0.0816$\pm$0.0006 (0.081$\pm$0.001 when including stellar mass and redshift) when all objects are included, and decreases to 0.069$\pm$0.0006 (0.068$\pm$0.0004) when excluding the 2.4\% (2.2\%) of outliers, defined as those objects for which the difference between spectroscopic and photometric metallicity exceeds 0.2 dex. For test set 2, the Root Mean Square Error is 0.09$\pm$0.005 (0.09$\pm$0.003 when including stellar mass and redshift)  when all objects are included and 0.069$\pm$0.002 (0.069$\pm$0.002) when the outliers are excluded, and the fraction of outliers is 4.2\% (3.9\%). The uncertainties quoted here are derived as the standard deviation from 10-fold cross validation performed on 80-20\% training/test set splits. The negligible difference in the results obtained by using photometry only and by also including derived quantities is a testimony to the power of machine learning algorithms, which are able to pick up information with limited guidance. However, it is interesting to notice that this quality is inherent to more sophisticated algorithms, such as Extra Randomized Trees. If a simpler algorithm like the Ridge regression is used, the performance of the algorithm on test set 1 is about 20\% worse when stellar mass and photometric redshift are not included as features (RMSE = 0.11 $\pm$ 0.02 vs 0.087 $\pm$ 0.002 for test set 1 and RMSE = 0.102 $\pm$ 0.006 vs 0.095 $\pm$ 0.005 for test set 2). This confirms the importance of investing time in selecting the best-performing algorithm for a given problem and data set. More performance metrics are summarized in Table \ref{Tab:BestResults} and in Fig. \ref{Fig:BestResults}.

\begin{table*}
\begin{center}
\resizebox{\linewidth}{!}{
\begin{tabular}{|c|cccccc|}
\hline
 &&& & &  & \\
Data set & Number of objects & Average/Median r magnitude  & RMSE (all objects) & RMSE (no outliers) & OLF & r$_2$ score\\
  & in training/test set & &  & &  & \\
\hline
&&&&&& \\
0.09 $<$ z $<$ 0.12 & 20253/5063 & 14.2/17.5 &  0.081 $\pm$ 0.001 & 0.068 $\pm$ 4e-4 & 0.022 $\pm$ 0.002 & 0.59 $\pm$ 0.007 \\
&&&&&& \\
&&& (0.0816 $\pm$ 6e-4) & (0.069 $\pm$ 8e-4) & (0.024 $\pm$ 0.006) & (0.57 $\pm$ 0.006) \\
&&&&&& \\
\hline
&&&&&& \\
0.2 $<$ z $<$ 0.25 & 2440/610 & 18.3/18.3   & 0.09 $\pm$ 0.003 & 0.069 $\pm$ 0.002 & 0.039 $\pm$ 0.007 & 0.76 $\pm$ 0.02 \\
&&&&&& \\
&&& (0.09 $\pm$ 0.005) & (0.069 $\pm$ 0.002) & (0.042 $\pm$ 0.008) & (0.73 $\pm$ 0.02) \\ 
\hline
\end{tabular}}
\end{center}
\caption{Results (root mean square error, outlier fraction, and r$_2$ score) for our best algorithm, Extremely Randomized Trees, for our two sample data sets. Numbers in parentheses on the second line correspond to results obtained only using photometry information (magnitudes and colors), without including derived quantities such as stellar mass and photometric redshift (although the latter was still used for object selection since the sample data sets are exactly the same in the two cases).
Values and uncertainties are calculated as the average and standard deviation of 10 random realizations of training/test sets. We are able to measure metallicity to within 0.1 dex for 84/82\% of the objects respectively, and only 2.2/3.9\% of objects have photometric metallicities that differ by more than 0.2 dex from the spectroscopically measured value. }
\label{Tab:BestResults}
\end{table*}
\vspace{1cm}

\subsection{Error estimation}
\label{subsec:error}

We wanted to estimate what fraction of the RMSE observed in the two test sets is due to the experimental uncertainty in the data, and what fraction is a systematic error due to our imperfect ability to recover metallicity from the available photometry. To do so, for each of the test sets, we built 50 simulated catalogs in the following manner: 1. For each observed band, we replaced each data point with a value sampled from a Gaussian distribution with mean equal to the observed value and standard deviation equal to its photometric uncertainty; 2. We replaced the redshift value with a value sampled from a Gaussian distribution with mean equal to the observed value and standard deviation equal to the photometric redshift error listed in the catalog; 3. We replaced the mass value with a value sampled from a Gaussian distribution with mean equal to the reported mass measurement and standard deviation equal to half the difference between the 84th and 16th percentile values listed in the catalog. We ran our machine learning algorithm to predict metallicity values for each of these 50 catalogs, and we evaluated the average (over the number of objects in the catalog) standard deviation of the 50 metallicity estimations for each object. We obtained an average scatter of 0.012 and 0.018 respectively for test set 1 and test set 2, suggesting that the contribution of the experimental uncertainties to the overall error budget in metallicity estimation (0.081 and 0.09, including outliers) is modest. 

\subsection{Diagnostics}
\label{subsec:Diag}
\begin{figure*}
\begin{center}
\includegraphics[width=\linewidth]{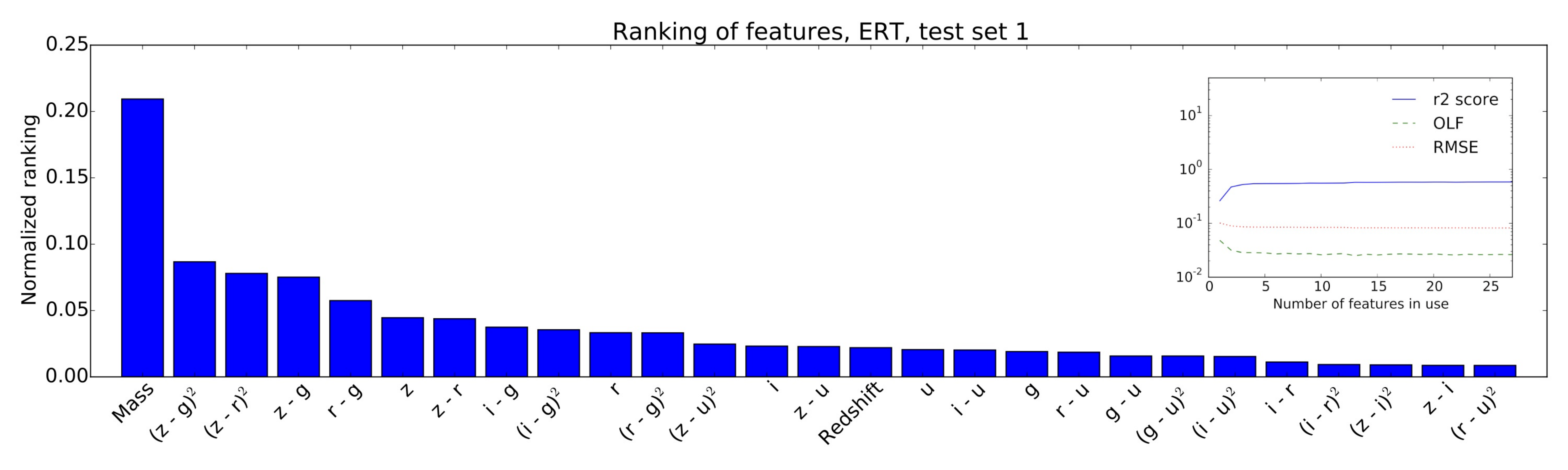} 
\includegraphics[width=\linewidth]{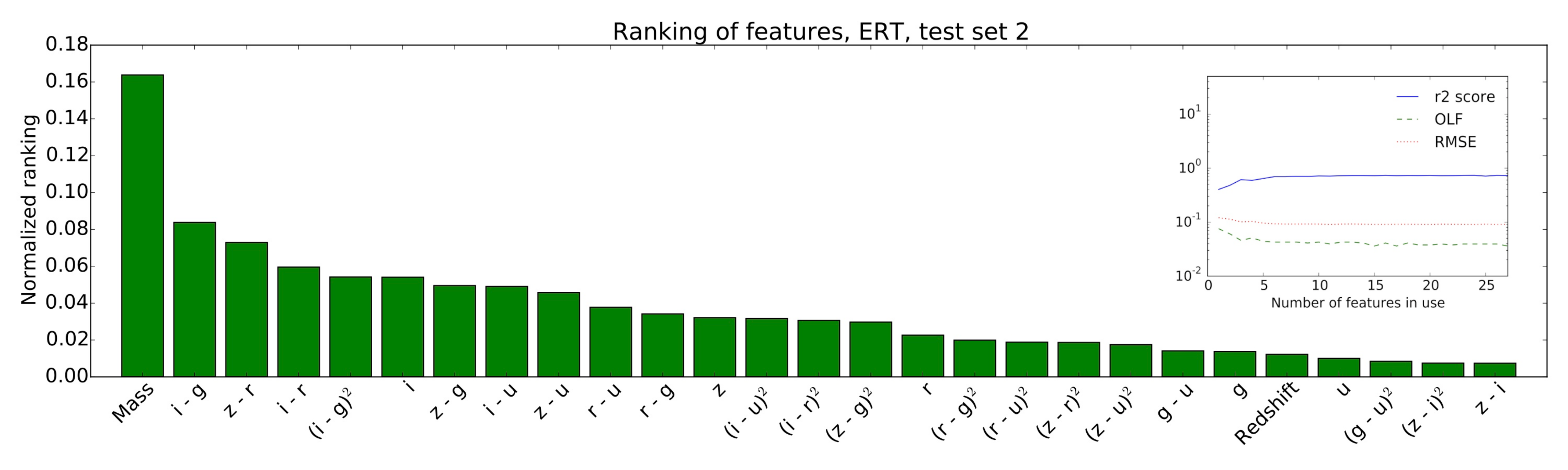} 

\caption{Feature ranking for the Extremely Random Trees estimator (ERT) the low-z (top) and high-z (bottom) test sets. The insets show how different performance diagnostics vary as features are progressively added in their ranked order. Mass is the most important feature in both cases, as a result of the well-known mass-metallicity relation; however, including information about luminosity, colors and square colors is essential to tighten the constraints on metallicity. The height of different columns in the histogram shows the contribution of each feature, with the sum normalized to one. These plots also show that different data sets might favor different features, suggesting that including all features and checking for over-fitting is preferable to a-priori feature selection.}
\label{fig:performance}
\end{center}
\end{figure*}

Despite their (undeserved) reputation of being a ``black box", machine learning methods offer a range of insights into the problems to which they are applied. For example, ``feature ranking" is a handy way of understanding which features are most important, and which ones might actually be harmful because they increase noise (variance) without reducing bias. For a tree-based algorithm such as ERT, the most important features will be used as decision nodes toward the top of the tree, and will contribute to the final prediction decision of a larger fraction of the input samples. The importance of each feature is calculated as the fraction of the input samples to which they contribute \citep{scikit-learn}. 
In Fig. \ref{fig:performance} we present the ranked features for the two test sets at median redshift of 0.103 and 0.218, using the ERT algorithm. We use the RMSE as the performance metric to rank features; the height of different columns in the histogram shows the contribution of each feature, with the sum normalized to one. Unsurprisingly, mass is the most relevant feature in the data set, but several other colors and squared colors contribute to reducing the bias and variance of the final estimate. In the figure, the insets show how some of the metrics perform on the test set, as a function of the number of features in use (sorted from most to least important). All curves have a monotonic behavior, indicating that while the bulk of the information is contained in the few most important features, there is some advantage (a 5-10\% difference in all metrics) in adding more features. The natural concern in this case is whether adding more features may lead to over-fitting, or lack of generalizability of our learning algorithm. However, on the one hand, ensemble methods such as ERTs are not prone to over-fitting since they build decision trees on random subsets of the features and of the data, and on the other, we do not observe any gap between the cross-validation scores and the test scores (in other words, between the performance on data that have and have not participated in the training process), so we can be confident that the quoted performance is accurate.

\subsection{Comparison with previous results}

\label{sec:comp}

The most relevant comparison of metallicity measurements from photometry is the work by \cite{Sanders2013}; hereafter S13. They recognize the possibility of extending the previously proposed luminosity-mass-metallicity relations by adding colors and their combinations in a fitting formula. Simple linear machine learning regression algorithms that use colors and their combinations as features will be equivalent to this approach; however, more sophisticated ML algorithms (such as Random Forests of trees, or SVMs with non-linear kernels) might be able to capture a more complicated relation between input features and predicted output (\ie metallicity). 
To test this hypothesis, we apply the exact same selection criteria to the SDSS data set as those applied by the authors of S13. The complete list can be found in their paper; the most significant cuts come from requiring that the objects are included in the SDSS main sample ($r$ mag $<$ 17.77), with redshift between 0.03 and 0.3, signal-to noise ratio in H$\alpha > 25$, and H$\alpha$/H$\beta$ flux $>$ 2.5. The most notable difference between the two methodologies is the fact that we don't need to apply any K-correction, since we use redshift as one of the features of our algorithm. S13 uses a slightly different indicator of performance, the scatter of the residuals vector (defined as the vector of differences between ``true" and estimated values). This is equivalent to the RMSE used in this paper as long as the mean of the residuals vector is zero. We have verified that the two estimates coincide up to the third significant digit for all of our test cases and we can compare the results directly.
From Fig. 2 of S13, we can see that the minimum scatter of residuals (as a function of their parameter $\alpha$) is 0.103; for the same sample, again using an 80-20\% training/test split and quoting the performance on the test set, we obtain a scatter of residuals of 0.0974 $\pm$ 0.0004, and 0.078 if the 4.8\% of outliers are excluded. It is interesting to note that if we apply less restrictive criteria, in particular we forfeit the cut based on the H$\alpha$ and H$\beta$ fluxes, the scatter of residuals actually decreases to 0.096, indicating that in this case a larger sample is more useful than cleaner data.

\begin{figure}
\includegraphics[width = \linewidth]{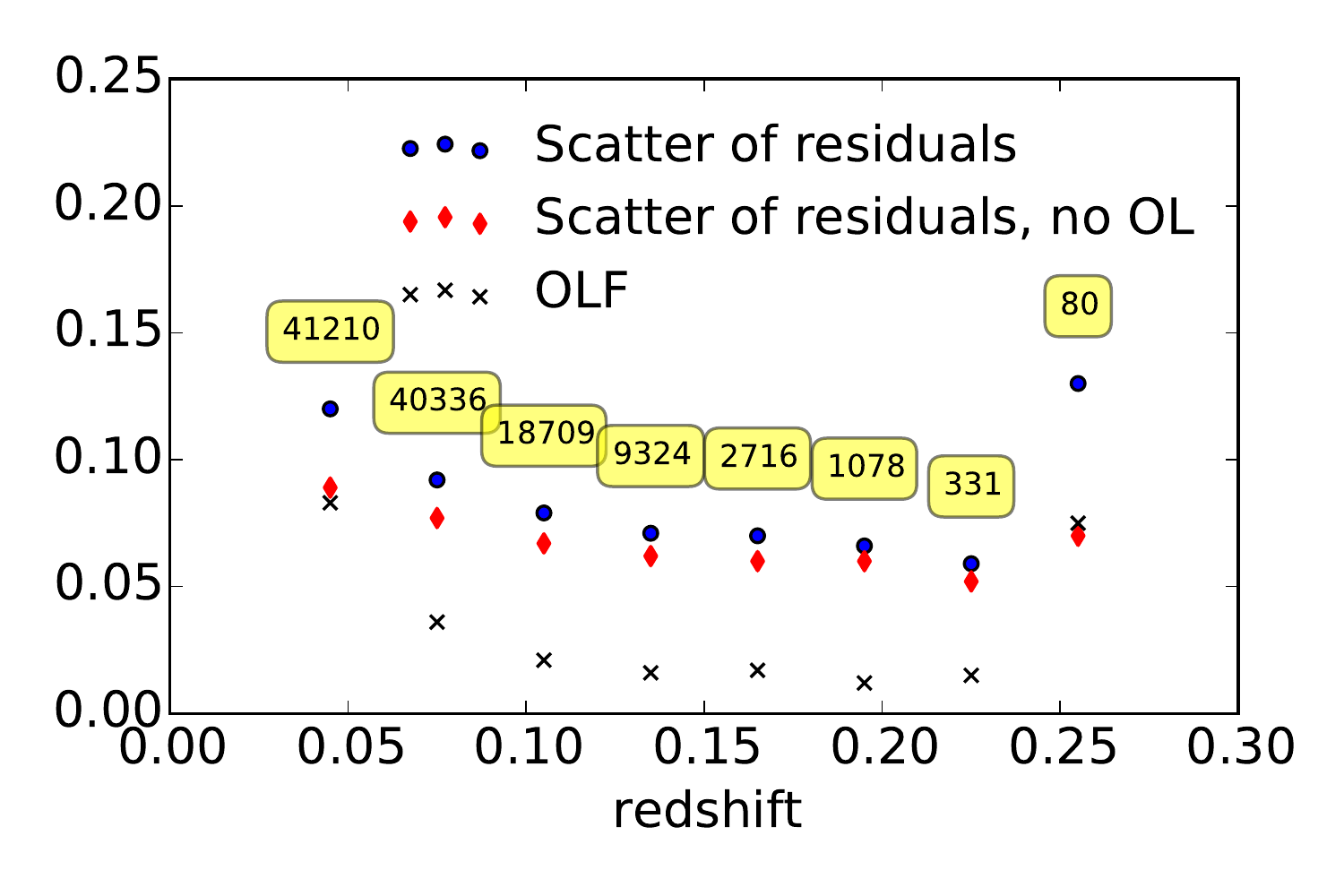}
\caption{Analysis of the objects in the S13 sample, divided in eight slices of uniform width $\delta z$ = 0.03 between $z =$ 0.03 and $z =$ 0.027. Sample sizes are indicated in the boxes. The objects in the first slice exhibit the highest fraction of outliers and the highest scatter of residuals, with the exception of the very small sample in the highest redshift slice.}
\label{fig:slices}
\end{figure}

It is also interesting to break down the performance of the algorithm by redshift, and number of objects. We divide the S13 sample in eight slices of width $\delta$z = 0.03 between 0.03 and 0.27, and train an ERT algorithm separately on each slice. The results, shown in Fig. \ref{fig:slices}, show that the average scatter of residuals in the results is dominated by the objects at redshifts 0.03 $< \,z \,<$ 0.06, which constitute a third of the sample and exhibit higher scatter (in fact, objects at $z <$ 0.07 were excluded in the mass-metallicity relation formulated by \citealt{Mannucci2010}). In all other slices, including those only populated by a few hundred objects, the number of outliers and the scatter of residuals are actually considerably lower, while they increase again in the last slice as a result of the excessively small (N = 80) sample size. 

\begin{table*}
\begin{center}
\resizebox{\linewidth}{!}{
\begin{tabular}{|c|lcccc|}
\hline
 &&& & &  \\
Test set & \quad \quad \quad EL list  & RMSE (all objects) & RMSE (no outliers) & OLF & r$_2$ score\\
  & & &  & &  \\
\hline
& None &  0.081 $\pm$ 0.001 & 0.068 $\pm$ 4e-4 & 0.022 $\pm$ 0.002 & 0.59 $\pm$ 0.007 \\
& NII & 0.073 $\pm$ 0.001 & 0.064 $\pm$ 4e-4 & 0.016 $\pm$ 0.002 & 0.65 $\pm$ 0.004 \\
& NII, OIII & 0.054 $\pm$ 0.001 & 0.048 $\pm$ 8e-4 & 0.006 $\pm$ 0.001 & 0.81 $\pm$ 0.007 \\
0.09 $<$ z $<$ 0.12  & NII, OIII, OII & 0.05 $\pm$ 0.001 & 0.044 $\pm$ 4e-4 & 0.005 $\pm$ 7e-4 & 0.84 $\pm$ 0.005 \\
& NII, OIII, OII, H$\beta$ & 0.048 $\pm$ 0.001 & 0.042 $\pm$ 4e-4 & 0.005 $\pm$ 6e-4 & 0.86 $\pm$ 0.004 \\
&  NII, OIII, OII,  H$\beta$, H$\alpha$ & 0.046 $\pm$ 0.001 & 0.042 $\pm$ 4e-4 & 0.0038 $\pm$ 3e-4 & 0.86 $\pm$ 0.002 \\
\hline
& None & 0.09 $\pm$ 0.003 & 0.069 $\pm$ 0.002 & 0.039 $\pm$ 0.007 & 0.76 $\pm$ 0.02 \\
&OII & 0.081 $\pm$ 0.004 & 0.063 $\pm$ 0.002 & 0.027 $\pm$ 0.005 & 0.79 $\pm$ 0.02 \\
&OII, NII & 0.061 $\pm$ 0.003 & 0.052 $\pm$ 0.002 & 0.013 $\pm$ 0.005 & 0.87 $\pm$ 0.008 \\
0.2 $<$ z $<$ 0.25 & OII, NII, OIII & 0.06 $\pm$ 0.002 & 0.05 $\pm$ 0.002 & 0.012 $\pm$ 0.004 & 0.87 $\pm$ 0.01 \\
& OII, NII, OIII, H$\beta$ & 0.057 $\pm$ 0.004 & 0.047 $\pm$ 0.002 & 0.01 $\pm$ 0.004 & 0.88 $\pm$ 0.008 \\
& OII, NII, OIII, H$\beta$, H$\alpha$& 0.058 $\pm$ 0.004 & 0.047 $\pm$ 0.001 & 0.01 $\pm$ 0.004 & 0.89 $\pm$ 0.01 \\
\hline
\end{tabular}}
\caption{Improvement of results when additional spectroscopic measurements of various emission lines are included. When using spectroscopic line flux measurements, we also assume that spectroscopic redshifts are available, although this makes very little difference for these data sets since the photometric redshifts have very small uncertainties.}
\label{tab:addEL}
\end{center}
\end{table*}

\begin{figure}
\begin{centering}

\end{centering}
\end{figure}

\begin{figure*}
\begin{center}
\includegraphics[width=0.49\linewidth]{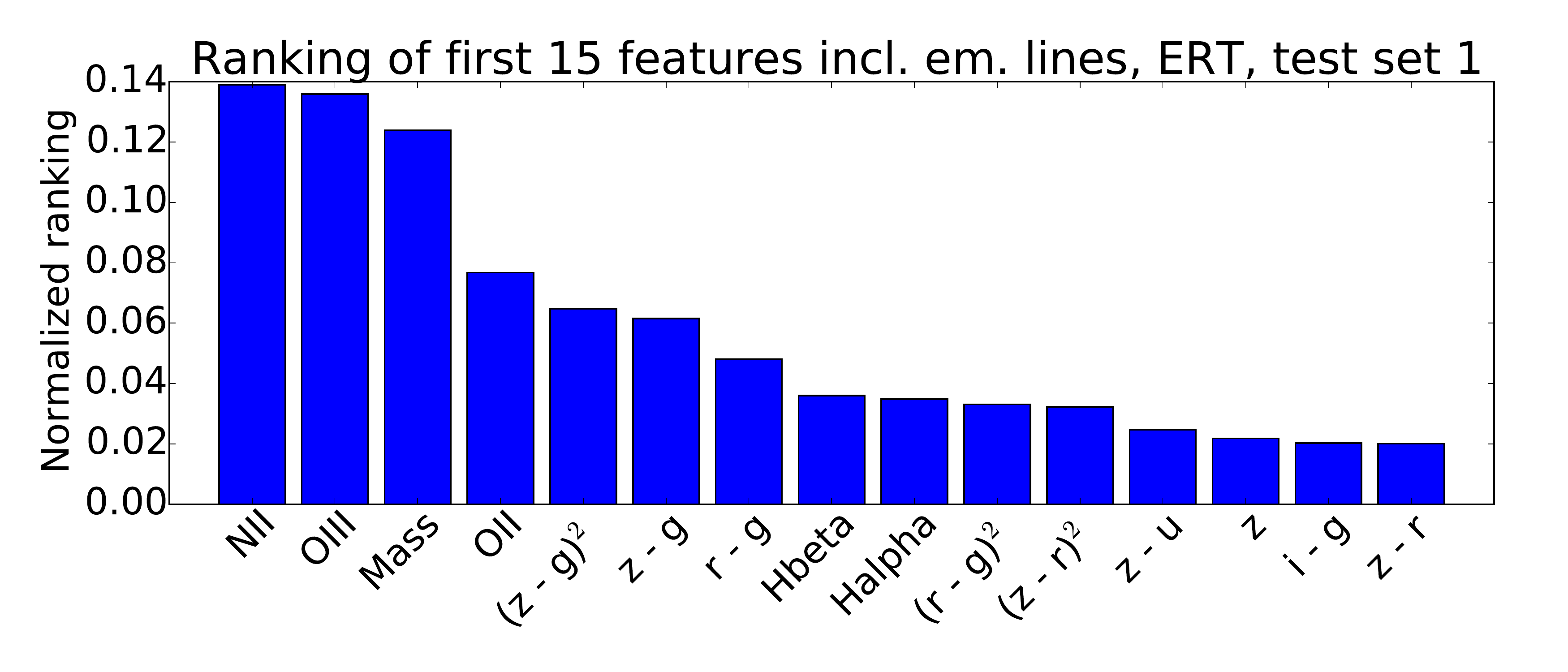} 
\includegraphics[width=0.49\linewidth]{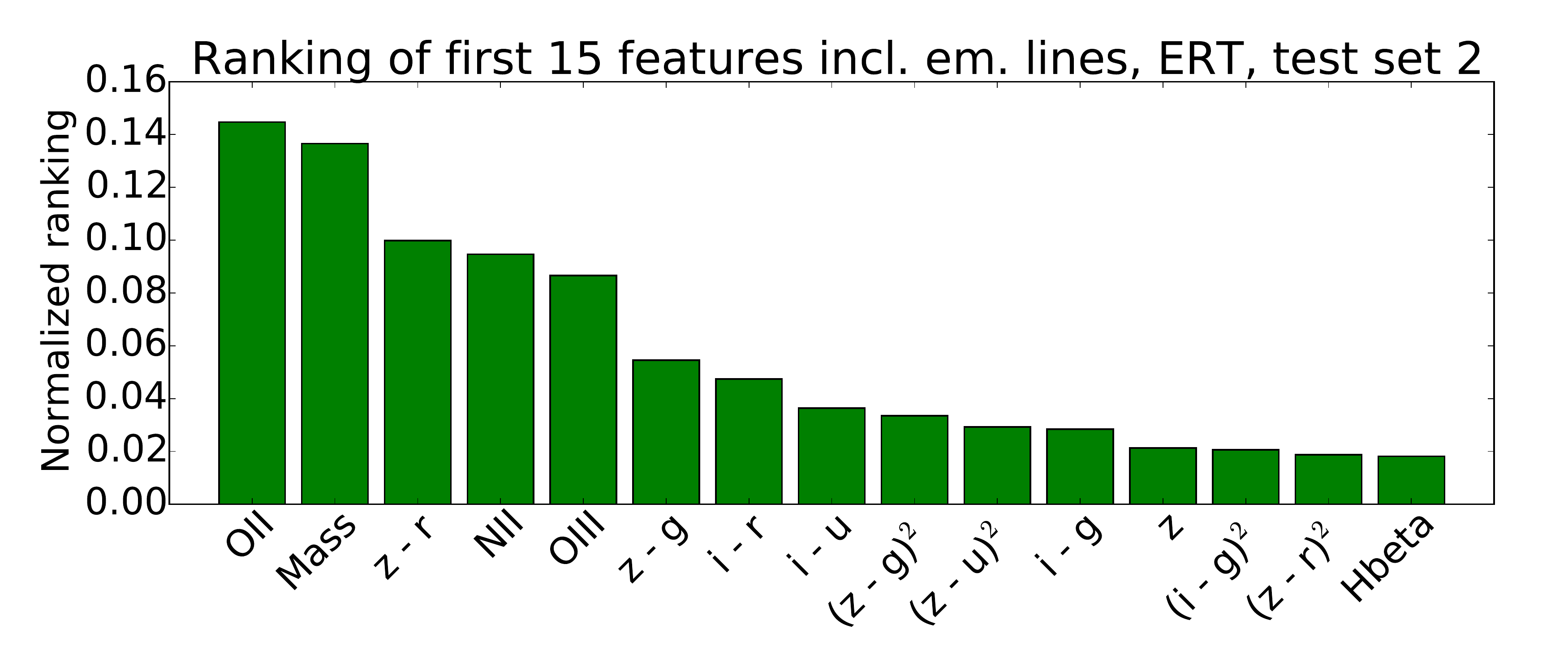} 
\caption{Feature ranking for the Extremely Randomized Trees estimator (ERT), once emission line measurements are included among the features of the algorithm. For both test sets, [OIII], [NII], and [OII] measurements are the most crucial, although the ranking differs between the two data sets. Furthermore, this analysis depends on the calibration of the spectroscopic metallicity indicator (here we use the values from \protect\citealt{Tremonti2004} as ground truth); using another reference system might lead to different results.}
\label{fig:first15}
\end{center}
\end{figure*}

\section{Combining spectroscopy and photometry}
\label{sec:spectrophoto}
Using machine learning algorithm allows for a seamless integration of spectroscopic and photometric data. This enables one to extract maximal information from available measurements of emission line fluxes, even if they are limited to some of the objects in the data set. Furthermore, the ``feature ranking" tool can be used to understand   
which emission line measurements are most helpful in constraining metallicity, and therefore help in planning follow-up spectroscopic campaigns. In this section, we add to our data set five additional emission line measurements that are available for the SDSS catalog: [OII] (doublet at 3726 and 3729 \AA), [OIII] (doublet at 4959 and 5007 \AA), [NII] (doublet at 6548 and 6584 \AA), H$\alpha$ at 6563 \AA, and H$\beta$ at 4861 \AA.

Unsurprisingly, adding all the five emission lines has a transformative impact on the ability to measure metallicity, with a reduction in the root mean square error of 40-50\% when all lines are included. By ranking the features in order of importance, as shown in Fig. \ref{fig:first15}, we observed that for both sample data sets, measurements of [OIII], [NII] and [OII] emission line fluxes were the most effective in increasing the accuracy of the metallicity measurement, and accounted for 90\% of the total improvement, although there were differences in the rankings between the two data sets. We estimated the impact of each of the five emission lines by adding them, one at a time according to their ranking, to our baseline data sets; the complete results are shown in Table \ref{tab:addEL}. When using spectroscopic line flux measurements, we also assume that spectroscopic redshifts are available, although this makes very little difference for these data sets since the photometric redshifts reported in the SDSS catalog have very small uncertainties. We note that the highest gain is obtained when adding a second emission line; for example, the RMSE for the first data set improves from 0.081 to 0.073 when adding the measurement of the [NII] line flux, and from 0.073 to 0.054 when adding the measurement of the [OIII] line flux. This confirms the well-known results that line ratios are more effective tracers of metallicity than single emission lines.

\section{Future applications: small data sets, LSST data, and higher redshifts}
\label{sec:applications}

We conclude our analysis by examining two issues: the applicability of this method to smaller data sets, and the perspectives for measuring metallicity with LSST data.

\begin{figure}
\includegraphics[width = \linewidth]{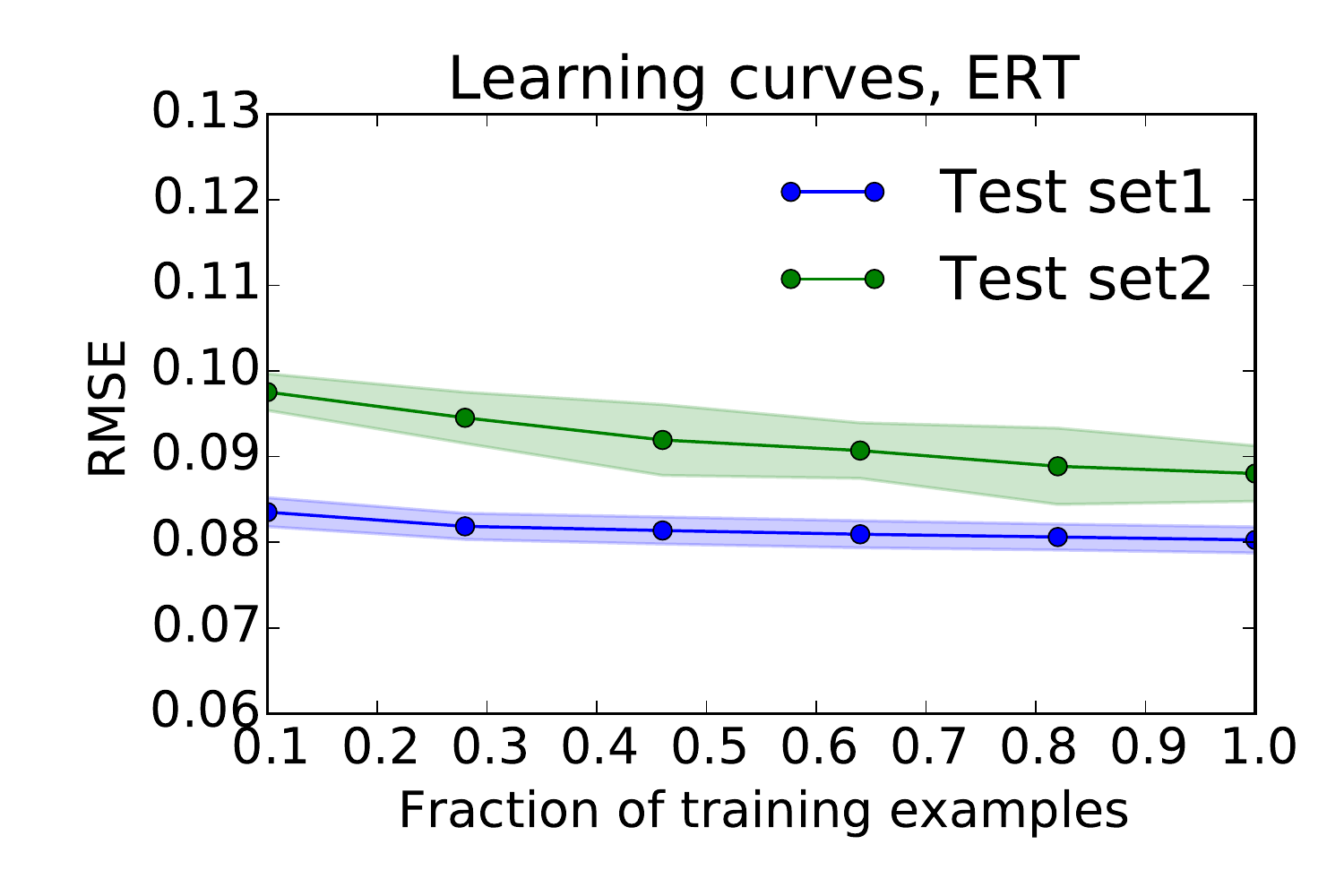}
\caption{Learning curves showing how the RMSE varies as a function of the size of the training set. For the first data set, the slope of the segments is essentially zero, indicating that collecting a larger training set would not be beneficial.  For the second, much smaller data set, the slope is slightly negative, indicating that collecting a larger training set is likely to improve the precision of the metallicity estimation. However, even having a training set only 20\% as large would only result in a performance lower by 5\% (RMSE varying from 0.09 to 0.095.)}
\label{fig:LearningCurves}
\vspace{0.5cm}
\end{figure}

In Fig. \ref{fig:LearningCurves} we plot the so-called ``learning curves" of our ERT algorithm. These diagrams show how the performance metrics (in this case, the RMSE of ground truth versus prediction) change as a function of the number of objects in the training set. To allow a fair comparison of the two sample data sets, we actually refer to the fraction of objects in the training set with respect to the full training set used in the previous sections. We can see that for the first data set, the slope at the far right of the plot is essentially zero, meaning that collecting new samples with spectroscopic metallicity would not improve the photometric metallicity determination. On the other hand, for the smaller data set 2, the slope is still negative at the far end of the curve and having more training examples would be beneficial. However, the gradient is small, and even for the second data set, having spectroscopic metallicity measurements for only 20\% of the objects in sample data set 2 - corresponding $\sim$ 670 objects - would only cause a few \% degradation in the results, with the RMSE varying from 0.09 to 0.095. This result is also aligned with what we found in the previous section when we explained the performance on the ERT algorithm on different slices of data with varying number of objects.

As a last step, we are interested in predicting whether the greater depth and additional waveband coverage provided by the LSST survey will result in a sizable improvement in the measurements of metallicity from photometry. Ideally, one could use a realistic simulated data set and apply the ML algorithm we devised to the simulated data. However, because modeling SEDs accurately as a function of metallicity is difficult, for the reasons described in the introduction, this procedure seriously underestimates the expected RMSE of ground truth versus predicted values. Therefore, we adopt a two-step approach that gives us the expected improvement relative to the current results, using our sample data set 2 as a reference. First, we use the nominal LSST uncertainties in the completed main survey (5-$\sigma$ limiting magnitudes of 26.3, 27.5, 27.7, 27.0, 26.2 and 24.9 respectively in bands u, g, r, i, z, and y, from \citealt{LSSTSciBook}), as opposed to the uncertainties from the SDSS catalog, to calculate the scatter due purely to photometric error, as described in Sec. \ref{subsec:error}. This step gives us an estimate of the impact of having deeper photometry. Second, we build a simulated galaxy catalog by running our SED fitting code, SpeedyMC \citep{GalMC, SpeedyMC}, on all the galaxies from sample data set 2, generating the best-fitting model SEDs, convolving them with the LSST filter transmission curves, and adding the appropriate photometric scatter to the simulated data points in each band. After obtaining the five- and six-band simulated catalogs, we run our ML algorithm on both of them, and we use the {\it relative} improvement in the RMSE to quantify the improvement due to the addition of the y-band.
Using the LSST photometric uncertainties reduced the RMSE due to photometric error by a factor of three for test set 1 (RMSE due to photometry decreasing from 0.013 to 0.004) and by a factor of two and a half for test set 2 (RMSE due to photometry decreasing from 0.018 to 0.008). The addition of a sixth photometric band was modestly helpful, presumably because observations in y-band in LSST will be considerably shallower than in the other bands. Overall, the projected improvement in the RMSE of truth-vs-prediction for these two sample data sets was about 5\%. However, this estimate doesn't take into account the improvement due to better measurements of photometric redshifts and stellar masses that will be available through LSST data, or the potential problems caused by wrong photometric redshifts, whose effect was however shown to be very minor in Sec. \ref{Sec:DataCleaning}. 

Perhaps more significantly, LSST data will enable similar quality measurements for galaxy samples several magnitudes deeper than the ones considered there, provided that a similar-depth training set with spectroscopic metallicity estimates is available. To quantify this effect, we have run our metallicity recovery algorithm on ``simulated" data sets created by artificially dimming the objects in our reference data sets, preserving the galaxy colors and the mass-luminosity-color-metallicity relation, and using the LSST projected depths in each band to estimate the S/N. We found that in both cases, the scatter in the measurement of metallicity due to the photometric error remained sub-dominant (contributing less than 20\% to the total scatter) for samples up to eight magnitudes deeper than the ones considered in this paper. 

We should also emphasize that the use of LSST data will enable a modest increase in the redshift range of applicability of our method. To obtain a rough estimate of the performance of LSST for data sets at a median redshift of $z \sim 0.3$, we used the fact that the rest-frame coverage of the six-bands LSST survey will be approximately equivalent to the coverage of the current SDSS survey in the u, g, r, i bands. We found that the anticipated metallicity measurements are comparable (with a 3-4\% increase in the RMSE) to those achieved for our two reference data sets. There are other very promising avenues for applying this method to significantly higher redshift samples, for example by using the synergy between the recently released spectroscopic data from 3D-HST survey \citep{3DHST2015} and the multi-wavelength photometric catalog of the CANDELS survey \citep{CANDELS}; we plan to pursue this project in a subsequent paper.

\section{Conclusions}
\label{sec:conclusions}

We have presented a novel method to determine gas-phase metallicity from photometry using supervised machine learning algorithms. Using the SDSS photometric catalog and the spectroscopically derived estimates of metallicity from \cite{Tremonti2004} as ground truth, we have built and optimized several estimators for two sample data sets at different redshifts and limiting magnitudes; we have trained the algorithms using multi-fold cross validation to avoid over fitting, and reported the results obtained on two test sets that were never part of the fitting/optimization procedure. 

Our main conclusions are the following:
\begin{itemize}
\item Among the algorithms that we considered, the best-performing ones were ensemble methods such as Random Forests and Extremely Randomized Trees, and Support Vector Machines. Extremely Randomized Trees proved to be a good all-purpose estimator, performing nearly to optimal in all cases with a CPU footprint 2-3 times lower than the other two methods.

\item When a sophisticated algorithm such as ERT is employed (as opposed to, e.g., linear regression), using only measured quantities (magnitudes and colors) as features provides comparably good results to the case where derived quantities such as stellar mass and photometric redshift are also included.

\item Supervised machine learning techniques allow one to measure gas-phase metallicity from five-band photometry with a typical uncertainty of 0.08-0.09 dex when all objects are considered. The number of outliers, defined as objects for which the predicted value differs front he ground truth by more than 0.2 dex, is very limited (2-4\% of the total sample); once these objects are excluded, the typical uncertainty decreases to less than 0.07 dex. This is a 3-fold reduction compared to the estimates of stellar metallicity from SED fitting for samples with higher depth and broader wavelength coverage (\eg \citealt{Dye2008,Pacifici2013}).

\item Our technique leads to improved results ($\sim$ 10\% lower scatter) over previously proposed methods that used fitting formulas of combinations of colors in addition to luminosity and mass to measure metallicity \citep{Sanders2013}. Furthermore, the technique is very forgiving, so that data cleaning or a careful selection of objects in the sample is unnecessary (and in fact, some times detrimental since ML algorithms are able to pick up useful features even in noisy data). In particular, K-corrections are not necessary once redshift is employed as one of the features of the ML algorithm.
 
\item This method enables metallicity measurements to within 0.1 dex even for small training sets comprised of a few hundreds of objects; in other words, the amount of spectroscopic measurements of metallicity necessary to build a solid metallicity estimator is limited. 
 
\item This technique allows one to easily combine photometric measurements with other spectroscopic measurements, such as measurements of one or more emission lines, even in the case of sparse data sets, when these measurements are only available for a fraction of the objects in the sample. 

\item An important caveat is that in order for this technique to be reliable, the spectroscopic sample used as a training set should be a fair, unbiased representation of the photometric sample to which the method is applied; for example, one should consider galaxies with similar stellar populations and at similar redshift.
 
\end{itemize}

In the hope that our work might be useful to others and to promote transparency in data analysis in science, we make available a Python notebook containing all the routines to reproduce the results in this paper, and to apply our method to other data sets, together with the data files we assembled for this project. They can be found at https://github.com/vacquaviva/Metallicity\_Estimation.

\section*{Acknowledgments}

It is a pleasure to thank Jarle Brinchmann for his help in navigating SDSS data, Lynne Jones for pointing us to the LSST filter transmission curves, 
Nathan Sanders for kindly sharing the list of objects used in his paper, Eric Gawiser, Maryam Modjaz and Samir Salim for useful comments on the manuscript, the participants of the ``Modeling galaxies through cosmic times" conference (IoA, Cambridge, UK, 09/15) for a lively discussion and useful feedback, and the anonymous referee for several suggestions that helped us improve the paper.


\vspace{1cm}

\begin{thebibliography}{}
\makeatletter
\relax
\def\mn@urlcharsother{\let\do\@makeother \do\$\do\&\do\#\do\^\do\_\do\%\do\~}
\def\mn@doi{\begingroup\mn@urlcharsother \@ifnextchar [ {\mn@doi@}
  {\mn@doi@[]}}
\def\mn@doi@[#1]#2{\def\@tempa{#1}\ifx\@tempa\@empty \href
  {http://dx.doi.org/#2} {doi:#2}\else \href {http://dx.doi.org/#2} {#1}\fi
  \endgroup}
\def\mn@eprint#1#2{\mn@eprint@#1:#2::\@nil}
\def\mn@eprint@arXiv#1{\href {http://arxiv.org/abs/#1} {{\tt arXiv:#1}}}
\def\mn@eprint@dblp#1{\href {http://dblp.uni-trier.de/rec/bibtex/#1.xml}
  {dblp:#1}}
\def\mn@eprint@#1:#2:#3:#4\@nil{\def\@tempa {#1}\def\@tempb {#2}\def\@tempc
  {#3}\ifx \@tempc \@empty \let \@tempc \@tempb \let \@tempb \@tempa \fi \ifx
  \@tempb \@empty \def\@tempb {arXiv}\fi \@ifundefined
  {mn@eprint@\@tempb}{\@tempb:\@tempc}{\expandafter \expandafter \csname
  mn@eprint@\@tempb\endcsname \expandafter{\@tempc}}}

\bibitem[\protect\citeauthoryear{{Acquaviva}, {Gawiser}  \&
  {Guaita}}{{Acquaviva} et~al.}{2011a}]{SpeedyMC}
{Acquaviva} V.,  {Gawiser} E.,   {Guaita} L.,  2011a,  \mn@doi [Proceedings of the IAU
  symposium ``The SED of galaxies", Preston, UK]{10.1017/S1743921312008691}, \href{
  http://adsabs.harvard.edu/abs/2012IAUS..284...42A} {284, 42}

\bibitem[\protect\citeauthoryear{{Acquaviva}, {Gawiser}  \&
  {Guaita}}{{Acquaviva} et~al.}{2011b}]{GalMC}
{Acquaviva} V.,  {Gawiser} E.,   {Guaita} L.,  2011b, \mn@doi [\apj]
  {10.1088/0004-637X/737/2/47}, \href
  {http://adsabs.harvard.edu/abs/2011ApJ...737...47A} {737, 47}

\bibitem[\protect\citeauthoryear{Caruana \& Niculescu-Mizil}{Caruana \&
  Niculescu-Mizil}{2005}]{Caruana05anempirical}
Caruana R.,  Niculescu-Mizil A.,  2005, in In Proc. 23 rd Intl. Conf. Machine
  learning (ICMLÍ06). pp 161--168

\bibitem[\protect\citeauthoryear{{Dav{\'e}}, {Finlator}  \&
  {Oppenheimer}}{{Dav{\'e}} et~al.}{2012}]{Dave2012}
{Dav{\'e}} R.,  {Finlator} K.,   {Oppenheimer} B.~D.,  2012, \mn@doi [\mnras]
  {10.1111/j.1365-2966.2011.20148.x}, \href
  {http://adsabs.harvard.edu/abs/2012MNRAS.421...98D} {421, 98}

\bibitem[\protect\citeauthoryear{{Dye}}{{Dye}}{2008}]{Dye2008}
{Dye} S.,  2008, \mn@doi [\mnras] {10.1111/j.1365-2966.2008.13639.x}, \href
  {http://adsabs.harvard.edu/abs/2008MNRAS.389.1293D} {389, 1293}

\bibitem[{{Grogin} {et~al.}(2011)}]{CANDELS}
{Grogin}, N.~A. {et~al.} 2011, \mn@doi[\apjs] {10.1088/0067-0049/197/2/35}, \href{http://adsabs.harvard.edu/abs/2011ApJS..197...35G} {197, 35}

\bibitem[\protect\citeauthoryear{{LSST Science Collaborations}}{{LSST Science
  Collaborations}}{2009}]{LSSTSciBook}
{LSST Science Collaborations} 2009, preprint, \href
  {http://adsabs.harvard.edu/abs/2009arXiv0912.0201L} {} (\mn@eprint {arXiv}
  {0912.0201})

\bibitem[\protect\citeauthoryear{{Lara-L{\'o}pez} et~al.,}{{Lara-L{\'o}pez}
  et~al.}{2010}]{LaraLopez2010}
{Lara-L{\'o}pez} M.~A.,  et~al., 2010, \mn@doi [\aap]
  {10.1051/0004-6361/201014803}, \href
  {http://adsabs.harvard.edu/abs/2010A\%26A...521L..53L} {521, L53}

\bibitem[\protect\citeauthoryear{{Mannucci}, {Cresci}, {Maiolino}, {Marconi}
  \& {Gnerucci}}{{Mannucci} et~al.}{2010}]{Mannucci2010}
{Mannucci} F.,  {Cresci} G.,  {Maiolino} R.,  {Marconi} A.,   {Gnerucci} A.,
  2010, \mn@doi [\mnras] {10.1111/j.1365-2966.2010.17291.x}, \href
  {http://adsabs.harvard.edu/abs/2010MNRAS.408.2115M} {408, 2115}
  
  \bibitem[\protect\citeauthoryear{{Momcheva} et~al.}{2015}]{3DHST2015}
{Momcheva} I. G,   et~al., 2015, preprint,  \href
  {http://adsabs.harvard.edu/abs/2015arXiv151002106M} {} (\mn@eprint {arXiv}
  {1510.02106})

\bibitem[\protect\citeauthoryear{{Pacifici}, {Charlot}, {Blaizot}  \&
  {Brinchmann}}{{Pacifici} et~al.}{2012}]{Pacifici2012}
{Pacifici} C.,  {Charlot} S.,  {Blaizot} J.,   {Brinchmann} J.,  2012, \mn@doi
  [\mnras] {10.1111/j.1365-2966.2012.20431.x}, \href
  {http://adsabs.harvard.edu/abs/2012MNRAS.421.2002P} {421, 2002}

\bibitem[\protect\citeauthoryear{{Pacifici}, {Kassin}, {Weiner}, {Charlot}  \&
  {Gardner}}{{Pacifici} et~al.}{2013}]{Pacifici2013}
{Pacifici} C.,  {Kassin} S.~A.,  {Weiner} B.,  {Charlot} S.,   {Gardner} J.~P.,
   2013, \mn@doi [\apjl] {10.1088/2041-8205/762/1/L15}, \href
  {http://adsabs.harvard.edu/abs/2013ApJ...762L..15P} {762, L15}

\bibitem[\protect\citeauthoryear{Pedregosa et~al.,}{Pedregosa
  et~al.}{2011}]{scikit-learn}
Pedregosa F.,  et~al., 2011, Journal of Machine Learning Research, 12, 2825

\bibitem[\protect\citeauthoryear{{Sanders}, {Levesque}  \&
  {Soderberg}}{{Sanders} et~al.}{2013}]{Sanders2013}
{Sanders} N.~E.,  {Levesque} E.~M.,   {Soderberg} A.~M.,  2013, \mn@doi [\apj]
  {10.1088/0004-637X/775/2/125}, \href
  {http://adsabs.harvard.edu/abs/2013ApJ...775..125S} {775, 125}

\bibitem[\protect\citeauthoryear{{Schlegel}, {Finkbeiner}  \&
  {Davis}}{{Schlegel} et~al.}{1998}]{SFD98}
{Schlegel} D.~J.,  {Finkbeiner} D.~P.,   {Davis} M.,  1998, \mn@doi [\apj] {10.1086/305772}, \href{http://adsabs.harvard.edu/abs/1998ApJ...500..525S} {500, 525}

\bibitem[\protect\citeauthoryear{{Tremonti} et~al.,}{{Tremonti}
  et~al.}{2004}]{Tremonti2004}
{Tremonti} C.~A.,  et~al., 2004, \mn@doi [\apj] {10.1086/423264}, \href
  {http://adsabs.harvard.edu/abs/2004ApJ...613..898T} {613, 898}

\bibitem[\protect\citeauthoryear{{de los Reyes} et~al.,}{{de los Reyes}
  et~al.}{2014}]{DeLosReyes2014}
{de los Reyes} M.~A.,  et~al., 2014, \mn@doi [\aj] {10.1088/0004-6256/149/2/79}, \href
  {http://adsabs.harvard.edu/abs/2015AJ....149...79D} {149,79} 

\makeatother
\end{thebibliography}
\end{document}